\magnification=\magstep1
\baselineskip=20 true pt
\hsize=160 true mm
\vsize=200 true mm

\def\cent{\centerline}
\def\scri{{\cal I}}
\def\endproof{\nobreak\kern5pt\nobreak\vrule height4pt width4pt depth0pt}
\def\ucs{{\widetilde {\cal D}}}
\def\ucsprime{{\widetilde {\cal D}'}}
\def\scra{\scri_{\alpha}}
\def\scrb{\scri_{\beta}}


\cent{\bf The Cosmic Censor Forbids Naked Topology} \cent{\it G.J. Galloway}
\cent{Dept. of Mathematics and Computer Science, University of Miami}
\cent{Coral Gables, FL 33124, USA}
\cent{and}
\cent{\it E. Woolgar}
\cent{Dept. of Mathematics, University of Saskatchewan} \cent{106 Wiggins
Road, Saskatoon, SK, Canada S7N 5E6} \vskip 2 true cm
\cent{\bf Abstract}
\noindent
For any asymptotically flat spacetime with a suitable causal structure
obeying (a weak form of) Penrose's cosmic censorship conjecture and satisfying
conditions guaranteeing focusing of complete null geodesics, we prove that
active topological censorship holds. We do not assume global hyperbolicity,
and therefore make no use of Cauchy surfaces and their topology. Instead,
we replace this with two underlying assumptions concerning the causal
structure: that no compact set can signal to arbitrarily small
neighbourhoods of spatial infinity (``$i^0$-avoidance''), and that no future
incomplete null geodesic is visible from future null infinity. We show that
these and the focusing condition together imply that the
domain of outer communications is simply connected. Furthermore,
we prove lemmas which have as a consequence that if a future incomplete null
geodesic were visible from infinity, then given our
$i^0$-avoidance assumption, it would also be visible from points of
spacetime that can communicate with infinity, and so would signify a true
naked singularity. \par\vfil\eject


\bigskip
\cent{\bf I. Introduction}

\smallskip
\noindent
The active topological censorship theorem of Friedman, Schleich, and Witt 
[1] (hereinafter FSW) states that, under suitable conditions (including
asymptotic flatness and an energy condition), any causal curve beginning
and ending at infinity in a globally hyperbolic spacetime must be
deformable to a curve that is always near infinity, and so cannot ``probe
the topology of spacetime.''

Since the original FSW result, several key issues have been resolved. Among
them, the FSW result has been extended to prove simple connectedness of the
domain of outer communications [2] and a ``local'' form of the censorship
theorem has been established [3]. A counter-example to the so-called passive
topological censorship theorem has been given [4]. In addition, the
theorem has been applied to answer questions concerning the topology of
black hole horizons [5,6].

In this paper, we wish to clarify the role of a chief assumption of the FSW
theorem, namely global hyperbolicity. To gain a heuristic understanding of
this issue, consider that the FSW theorem is inspired by the singularity
theorems of Gannon [7], who showed that under physically reasonable
conditions, there are future-incomplete geodesics lying in the future
development of any non-simply connected partial Cauchy surface that is
``regular near infinity'' (a form of asymptotic flatness). Given the FSW
assumptions of global hyperbolicity and asymptotic flatness at null infinity,
then we expect these incomplete geodesics to pass behind horizons, so as
not to be visible from any point that can signal future null infinity.
Then one could interpret the FSW theorem as the statement that the horizons
arise early enough to also trap any causal curves which would otherwise
``traverse the topology'' (and connect up with a causal curve near infinity
to form a non-trivial loop).

The role played by horizons in this discussion suggests that the global
hyperbolicity assumption is being used to enforce an underlying assumption,
namely (weak) cosmic censorship. This has led us to seek a proof of the
active topological censorship theorem that does not assume global
hyperbolicity, but assumes instead that no naked singularities appear. The
purpose of this paper is to present such a theorem, replacing the global
hyperbolicity condition by the requirement that there are no
future-incomplete null geodesics ``visible'' (in a sense to be made
precise) from a point of the domain of outer communications.

However, global hyperbolicity also plays another role in the FSW theorem.
This theorem rests on a lemma which
holds only if causal curves
beginning in any fixed compact set in spacetime cannot arrive at $\scri^+$
at arbitrarily early times --- we term this ``$i^0$-avoidance.''
This in turn
follows from global hyperbolicity, though strictly this inference entails
extending the concept of global hyperbolicity to refer to causal curves
ending on $\scri$. We will need to use the $i^0$-avoidance property herein,
but since we make no global hyperbolicity assumption from which it can be
derived, we instead impose it {\it a priori}. It can be rephrased as a type
of causality condition imposed only at spatial infinity; specifically, it
is the condition of {\it causal continuity at} $i^0$.

In the next section, we prove our topological censorship theorem. We first
prove that a failure of topological censorship would result in an
incomplete null geodesic visible from a point of $\scri^+$. We then prove
that such a geodesic would necessarily also be visible from points within
the domain of outer communications. In the final section, we give some
remarks on the importance of the assumption of causal continuity at $i^0$
and on the geodesic focusing condition that we introduce below.


\bigskip
\cent{\bf II. Censorship Theorem}

\smallskip
\noindent
We will assume the boundary-at-infinity of an asymptotically flat spacetime
to be a null cone with vertex $i^0$ representing spatial infinity. We
denote it by $\scri=\scri^-\cup \{ i^0 \} \cup \scri^+$ and its domain of
outer communications by
$${\cal D}:=I^+(\scri^-)\cap I^-(\scri^+)\quad .$$ It is convenient to
extend a little beyond
$\scri$; then ${\cal D}$ is embedded in a larger manifold and
$\scri$ is its boundary. We
remark that in the more general case of a boundary which is a disjoint
union of cones, our
arguments apply to each cone and its associated domain of outer
communications individually. We
will also assume that there is a neighbourhood ${\cal N}$ of $\scri$ such
that ${\cal N}\cap
{\cal D}$ has topology $\Re\times\Sigma$ where the $\Re$ fibres are
timelike and the surfaces
$\Sigma$ are spacelike and have the topology of the region exterior to a
sphere in $\Re^3$ ---
note that this yields what are referred to as simply connected
neighbourhoods of infinity (one
of these surfaces may be completed by appending $i^0$ to it, the rest are
bounded by 2-sphere
cuts of $\scri$).

It is helpful to introduce the following definition.

\medskip
\noindent
{\bf Definition 1:} A future-inextendible causal curve $\zeta$ will be said to
be {\it visible} from a point $x$ iff it is contained in
${\overline {I^-(x)}}$ (the overhead bar indicates closure).

\medskip
\noindent
Although $\zeta$ may lie entirely outside $J^-(x)$, if it is visible
from $x$ in the sense of Definition~1 then $I^-(\zeta)\subseteq I^-(x)$;
{\it i.e.} every point of the {\it indecomposable past set} defined by
$\zeta$ can signal $x$. It is in this manner that we interpret the
term ``visible.'' This sense is used when discussing visibility of
singularities whose presence is indicated by future-incomplete null
geodesics.  This point of view is similar to that previously put
forth by Penrose [8], in which the version of cosmic censorship denoted 
by ``CC8'' is essentially that which we use as assumption ({\it iii}) in 
our first (and key) result:

\medskip
\noindent
{\bf Theorem 2:} {\sl Let ${\cal D}$ be the domain of outer communications
of an asymptotically flat spacetime ${\cal M}$ and let ${\cal H}^+:=
I^+(\scri^-)\cap \partial I^-(\scri^+)$ be the union of all future event
horizons that may be present. If

\item{({\it i})}{causal continuity holds at $i^0$,}
\item{({\it ii})}{every complete null geodesic $\gamma\subseteq{\cal D}
\cup{\cal H}^+$ possesses a pair of conjugate points, and}
\item{({\it iii})}{every future-inextendible null geodesic visible from
any point $q\in\scri^+$ is future-complete in ${\cal M}$,}

\noindent
then ${\cal D}$ is simply connected.
}

\medskip
\noindent
The following remarks discuss, in order, assumptions ({\it i--iii}).

\medskip
\noindent
{\bf Remark 2.1:}
The set-valued function $I^+$ is causally continuous at a point $x$ iff
for each compact set $K$ that does not meet ${\overline {I^+(x)}}$, there
is a neighbourhood $O\ni x$ such that, for any $y\in O$, $K$ does not meet
${\overline {I^+(y)}}$; dually for $I^-$. Causal continuity is said to hold
at $x$ if both $I^+$ and $I^-$ are causally continuous there.
At $x=i^0$, since no subset of spacetime meets ${\overline {I^-(i^0)}}
\cup {\overline {I^+(i^0)}}$, causal continuity
reduces to the statement that
for any compact set $K\subseteq{\cal D}$, then $i^0\notin {\overline
{I^-(K)}}\cup{\overline {I^+(K)}}$. It is this
phrasing which we employ in the proof.

\medskip
\noindent
{\bf Remark 2.2:} By standard theorems, complete null geodesics always have
a pair of conjugate points provided the so-called {\it null generic
condition}\footnote{$^1$}
{This is the condition that for each complete null geodesic, if $n^a$ is
tangent to the geodesic, then $n^an^bn_{[c}R_{d]ab[e}n_{f]}$ must be
non-zero somewhere along the geodesic.}
and a suitable energy condition hold. While the pointwise null energy
condition $R_{ab}n^a n^b\ge 0$ for all null $n^a$ suffices, so do the much
milder integrated energy conditions, such as the Averaged Null Energy
Condition [9] and the Borde condition [10].

\medskip
\noindent
{\bf Remark 2.3:} Any null geodesic beginning on $\scri^-$ that is visible
from a point of $\scri^+$ is necessarily contained in ${\cal D}\cup {\cal
H}^+$. If such a geodesic is future-complete in ${\cal M}$, it is also
future-complete in any open set containing ${\cal D}\cup {\cal H}^+$.

\medskip
\noindent
{\bf Proof:} By way of contradiction, we will assume that conditions ({\it
i}) and ({\it ii}) hold and that ${\cal D}$ is {\it not} simply connected,
and prove that ({\it iii}) cannot hold in these circumstances.

We begin by extending ${\cal D}$ somewhat into any black hole regions that
may be present.
Then ${\cal H}^+=I^+(\scri^-)\cap \partial I^-(\scri^+)$
is an achronal hypersurface in
$I^+(\scri^-)$ (viewed as a spacetime in its own right). Let $h_{ab}$ be a
Lorentzian metric on ${\cal M}$ strictly
narrower than the given metric $g_{ab}$ ({\it i.e.}, any causal vector of
$h_{ab}$ is timelike
in
$g_{ab}$). Then ${\cal H}^+$ is an edgeless acausal hypersurface --- {\it
i.e.}, a partial Cauchy
surface --- in the spacetime $(I^+(\scri^-),h_{ab})$. Let ${\cal U}$ be the
interior of the
domain of dependence of ${\cal H}^+$ in $(I^+(\scri^-),h_{ab})$. Finally,
let ${\cal D}' = {\cal
D}\cup {\cal U}$. Observe that by construction there is a deformation
retraction of ${\cal D}'$
onto
${\cal D} \cup {\cal H}^+$, whence $\pi_1({\cal D}') = \pi_1({\cal D} \cup
{\cal H}^+)$ $=\pi_1({\cal D})$,
where the second equality follows, for example, by the Seifert-Van Kampen
theorem.

We will employ the universal covering spacetimes and projections
$\pi:\ucs\to{\cal D}$ and $\pi':\ucsprime\to{\cal D}'$. By our
construction, we may embed $\ucs$ in $\ucs'$; then the region
$\ucs'\backslash\ucs$ is not in causal contact with future null infinity
(if it were otherwise, the causal curves mediating such contact would
project to curves which would causally connect a black hole in spacetime to
$\scri^+$).

The boundary-at-infinity of $\ucsprime$ will have at least two disjoint
components (we will extend $\pi'$ to this boundary). Let $S$ be a set
indexing these components. By a result of [2], there exist $\alpha,
\beta\in S$ (with $\alpha\neq\beta$) such that the distinct components
$\scra$ and $\scrb$ are joined by a causal curve, say beginning on
$\scra^-$ and ending at some point $Q\in\scrb^+$.\footnote{$^2$}
{The argument of [2] may be summarized as follows. By definition, every
point of ${\cal D}$ lies on some causal curve from $\scri^-$ to $\scri^+$.
By lifting these curves, we see that every point of $\ucs$ lies on a causal
curve from some $\scra^-$ to some $\scrb^+$ (where possibly
$\alpha=\beta$), so these causal curves cover $\ucs$. If every infinite
causal curve in $\ucs$ were to begin and end on the same component of its
$\scri$, then $\{I^+(\scra^-)\cap I^-(\scra^+)| \forall\alpha \in S\}$
would be a disjoint open cover for $\ucs$, which violates the fact that
$\ucs$ is connected, unless $S$ has only one element. Therefore if distinct
components of infinity exist for $\ucs$, then each such component must be
joined to some other by a causal curve. While this argument was made in
${\cal D}$ and its covering spacetime, since $\ucs\subseteq\ucsprime$ and
since their boundaries-at-infinity correspond, the result also applies to
$\ucsprime$.}

Now ${\overline {I^-(Q)}}$ does not contain $i^0_{\alpha}$. To see this,
consider any causal curve $\Lambda$ from $\scra^-$ to $Q\in\scrb^+$, and
project it back to
${\cal D}'\cup\scri$, obtaining a curve $\lambda$ that begins on $\scri^-$
and ends on
$\scri^+$. By joining a null generator of $\scri$ on to this curve, one can
form a closed curve,
which is necessarily a non-trivial loop. By the assumed asymptotic
topology, every loop in the
asymptotic region is trivial, so $\lambda$ must pass through the inner
boundary of the
asymptotic region. Fix an infinite timelike cylinder $C$ surrounding this
boundary --- clearly
$\lambda$ must meet this cylinder as well, as must any other curve
``threading the topology.'' Consider the spacelike surface $\Sigma_t$
(a leaf of the asymptotic topology $\Re\times\Sigma$) whose boundary includes
$q:=\pi'(Q)\in \scri^+$. The surface $S_t:=C\cap
\Sigma_t$ is a 2-sphere and, following any curve into the past from $q$, if
the curve meets $C$
it does so in the past of $S_t$. Since $S_t$ is compact, then by assumption
${\overline
{I^-(S_t)}}$ does not contain $i^0$, so no sequence of curves of this type,
considered prior to their final crossing of $C$, can approach $i^0$.
This implies that no sequence of causal curves in $\ucsprime$ from
$\scri^-_{\alpha}$ to $Q$ can
approach
$i^0_{\alpha}$, which is what we set out to show.

Since there is at least one causal curve from $\scra^-$ to $Q$ and since
$i^0_{\alpha}\notin {\overline {I^-(Q)}}$, then $\partial I^-(Q)$ must meet
$\scra^-$. Let $P$ lie in this intersection. From $P$, there will be a
future-null geodesic $\Gamma$ which
enters $\ucsprime$ (since $\Gamma$ cannot contain $i^0_{\alpha}$),
remains on $\partial I^-(Q)$, and does
not have a future endpoint other than $Q$.

Now $\Gamma$ cannot have conjugate points, since otherwise standard
results would imply that it could not lie on $\partial I^-(Q)$. Then its
projection $\gamma=:\pi'(\Gamma)$ cannot have conjugate points either.
But $\gamma$ enters ${\cal D}'$ (since $\Gamma$ enters $\ucsprime$)
and $\gamma \subseteq {\overline {I^-(q)}}$ (since
$\pi'({\overline {I^-(Q)}})\subseteq {\overline {I^-(q)}}$), so
$\gamma\subseteq {\cal D}\cup{\cal H}^+$. By assumption, every
complete null geodesic in ${\cal D}\cup{\cal H}^+$ has conjugate points, so
$\gamma$ must be incomplete, considered as a curve in ${\cal D}'$.

By Remark 2.3 this geodesic is incomplete in the spacetime ${\cal M}$.
Since it lies in ${\overline {I^-(q)}}$, it is visible from $q\in\scri^+$.
\hfill\endproof \medskip

We now seek to show that the incomplete geodesic $\gamma$ is visible not
only from a point of $\scri^+$ but also from a point of spacetime which can
signal $\scri^+$. To show this, we give a useful definition and then two
lemmata based upon it. Then our final censorship theorem will follow
immediately from these results. 

\medskip
\noindent
{\bf Definition 3:} We will say that a future-inextendible causal curve
$\gamma$ is {\it weakly visible} from a point $x$ iff for every $t$ in the
domain of $\gamma$ there is a point $x_t$ such that\footnote{$^3$}
{Here, and in what follows, we use {\it nets} rather than sequences. In the present
context, this is merely a convenience, allowing us to use the continuous parameter
$t$ as an index rather than having to introduce an integer $n$. The reader more
comfortable with sequences may here define an increasing sequence $t_n$ and replace
$\gamma(t)$ and $x_t$ by $\gamma(t_n)$ and $x_n$ in the present definition. Similar
modifications may then be made in what follows.}
\item{(i)}{$\gamma(t)\in {\overline {I^-(x_t)}}$, and} 
\item{(ii)}{the net of points $x_t$ has $x$ as an accumulation point.}

\medskip
If a future-inextendible causal curve $\gamma$
is visible from $x$, then it is weakly visible from $x$, but the converse
to this is not always true. However, we have the following lemma:

\medskip
\noindent
{\bf Lemma 4:} {\sl If a future-inextendible causal curve $\gamma$ is
weakly visible from $z$, then it is visible from all points $w\in I^+(z)$.}

\medskip
\noindent
{\bf Proof:} First we choose some $T$ in the domain of $\gamma$ and show
that $\gamma(T)\in {\overline {I^-(w)}}$. To do this, note that $I^-(w)$ is
a neighbourhood of $z$. By definition, there is a net of points $z_t$ with
accumulation point $z$ such that $\gamma(t)\in {\overline {I^-(z_t)}}$. It
follows that for any choice of $T$ in the domain of $\gamma$, and hence for
our choice, there is a $t>T$ such that $z_t\in I^-(w)$ and $\gamma(t) \in
{\overline {I^-(z_t)}}$, whence $\gamma(t)\in {\overline {I^-(w)}}$. But
then it is easy to show that $\gamma(T)\in {\overline {I^-(w)}}$, since
$\gamma(t)\in {\overline {I^-(w)}}$ and since $\gamma(T)\in
J^-(\gamma(t))$.

But we can repeat the argument (with the same $w$) for all choices of $T$
in the domain of $\gamma$, so $\gamma\subseteq {\overline {I^-(w)}}$.
\hfill\endproof

\medskip

In virtue of this lemma, we need only show that our incomplete geodesic is
weakly visible from some point of spacetime, whence it will also be {\it
visible} from some point. But by the next lemma, weak visibility from a
point of spacetime follows, given weak visibility from a point of $\scri^+$
and ``$i^0$-avoidance'' (our causal continuity assumption).

\medskip
\noindent
{\bf Lemma 5:} {\sl If spacetime contains a future-inextendible causal
curve $\gamma$ weakly visible from a point of $\scri^+$ and if causal
continuity holds at $i^0$, then $\gamma$ is weakly visible from some point
$w$ in the domain of outer communications.}

\medskip
\noindent
{\bf Proof:} Take $\gamma$ to be weakly visible from some $q\in\scri^+$.
Along the null generator of $\scri^+$ from $i^0$ to $q$, let $Q$ be the
infimum of all points from which $\gamma$ is weakly visible.

Note first that $Q\neq i^0$. To see this, let $K$ be a compact
neighbourhood about some point $\gamma(T)$ in the incomplete curve. Then
$I^+(K)$ will be a neighbourhood of $\gamma(t)$ for each $t>T$. Then by the
definition of weak visibility, for each $t>T$ there is a sequence of
timelike curves ending on $x_t$ ({\it cf.} Definition 3) whose past
endpoints converge on $\gamma(t)$, so these curves eventually all lie in
$I^+(K)$. Then for $t>T$, the $x_t$ themselves obviously lie in $I^+(K)$.
It follows that the accumulation points $x$ of the net of points $x_t$ are
contained in ${\overline {I^+(K)}}$. In turn, since the points $x$ that lie
along the given generator of $\scri^+$ themselves constitute a net in
${\overline {I^+(K)}}$, then ${\overline {I^+(K)}}$ also contains the
infimum $Q$ of the points $x$. But by the assumption of causal continuity
at $i^0$, ${\overline {I^+(K)}}$ does not contain $i^0$, since $K$ is
compact, whence $Q\neq i^0$.

Now let $\tau$ be the supremum in $\Re$ of the domain of $\gamma$, and
parametrize the null generator of $\scri^+$ through $Q$ so that the
parameter has value $\tau$ at $Q$ and decreases to the future. Denote the
points of the generator by $q(s)$. By assumption, $\gamma$ is weakly
visible from $q(s)$ for $s<\tau$, so for each $s<\tau$ there is a net of
points $x_t(s)$ accumulating at $q(s)$ and satisfying Definition 3. From
these nets, construct the new net of points $x_s(s)$, and note that it has
$Q$ as an accumulation point (incidentally, this construction proves that
$\gamma$ is weakly visible from $Q$), whence it contains subnets which
converge to $Q$; let $z_t$ belong to such a subnet.

Choose a neighbourhood $U\ni Q$, small enough so that the contact between
${\overline {I^-(Q,U)}}$ and $\scri^+$ is along one null
generator.\footnote{$^4$}
{While this is usually the case for all such $U$, it could fail if
causality violations occurred at $Q$ and if $U$ were big enough to wholely
contain the curves responsible.}
Surround $Q$ by a closed surface $S\subseteq U$ with interior $O$. Then the
$z_t$ are ``eventually all'' ---{\it i.e.} for $t>T$ --- inside $O$. Since
$\gamma(t)\subseteq {\overline {I^-(z_t)}}$, for each $t$ there is a
sequence of causal curves, each curve having future endpoint $z_t$, whose
past endpoints converge to $\gamma(t)$. For $t>T$, these curves all meet
and penetrate $S$, and some may even do so more than once, but there is
always a final penetration point along each curve as it enters $O$ for
the last time and proceeds to its endpoint at $z_t$. For each $t>T$, one
can construct such a ``sequence of (final) penetration points'' and it will
have an accumulation point on $S$ since $S$ is compact; label this point
$a(t)$ (if there are multiple accumulation points, choose any one).
Increasing $t$ within the domain of $\gamma$, we obtain a net of
accumulation points $a(t)$, and this net will have it's own accumulation
point $y\in S$ (again, if there are multiple accumulation points, choose
any one). Note that $\gamma$ is weakly visible from $y$.

The point $y$ must lie in ${\overline {I^-(Q,U)}}$ but cannot lie along
$\scri^+$ (since $Q$ is an infimum for the set of points on the given
generator of $\scri^+$ from which $\gamma$ is weakly visible). Therefore,
$y\in{\cal D}$.\hfill\endproof

\medskip
Finally, we state our ultimate result, whose proof now follows trivially
from the preceding results.

\medskip
\noindent
{\bf Theorem 6:} {\sl If an asymptotically flat spacetime with domain of
outer communications ${\cal D}$ obeys conditions ({\it i}) and ({\it ii})
of Theorem 2 and if no future-incomplete null geodesic is visible
({\it cf.} Definition 1) from any point $x\in{\cal D}$,
then ${\cal D}$ is simply connected.}

\medskip
\noindent
{\bf Proof:} By Theorem 2, a non-simply connected domain ${\cal D}$
could be compatible with assumptions ({\it i}) and ({\it ii}) of that
theorem only if there were to exist an incomplete null geodesic $\zeta$
visible --- and therefore {\it weakly} visible --- from a point
of $\scri^+$. But this would violate the new assumption of Theorem 6,
for by Lemma 5, $\zeta$ would be weakly visible from
some point $y\in{\cal D}$ and hence, by Lemma 4, visible from
any point $x\in I^+(y)$. Therefore, if the assumptions of Theorem 6 hold,
then ${\cal D}$ must be simply connected.\hfill\endproof


\bigskip
\cent{\bf Final Remarks}
\smallskip
\noindent
We have given an active topological censorship theorem in which the usual
assumption of ``extended'' global hyperbolicity is discarded in favour of
two conditions which would ordinarily follow from it under the
circumstances, namely cosmic censorship and ``$i^0$-avoidance,''
the latter expressed as a causal continuity condition at $i^0$.

A question that arises is whether one might eventually weaken or altogether
do away with this last condition. In the introductory section, we argued
that the causal continuity condition at $i^0$, sometimes in the guise of a
generalized global hyperbolicity condition, plays an important role in
previous proofs of topological censorship.
Moreover, there is some evidence that one cannot completely dispense with
conditions of this general sort. In particular, we refer to recent work of
Schein and Aichelburg [11], taken together with an example quoted in [12]
and attributed there to Bardeen. Ref.~[11] gives a traversable wormhole
solution satisfying energy conditions. This solution has
incomplete geodesics and violates the $i^0$-avoidance assumption; indeed,
through every point of the asymptotic region there is a closed timelike
curve. However, the incomplete geodesics arise from singularities of the
sort familiar from the extended Reissner-Nordstr\"om solution, and
according to the Bardeen example it should be possible to remove them. If
so, the resulting spacetime
(or, rather, a perturbation thereof, which appears necessary to enforce
the null generic condition and concomitant focusing lemma) would provide
an example of a spacetime that obeys all our conditions except the causal
continuity condition at $i^0$, and in which topological censorship
violations arise as a result.

We close with a remark concerning our requirement that complete geodesics
should have conjugate points. Remark 2.2 notes that this follows from
integrated energy conditions and the null generic condition. While the FSW
theorem also follows from these assumptions, it follows as well from an
integrated energy condition on null geodesics that are complete in only the
future direction [2]. In this form of the FSW theorem, the generic
condition is not needed. Then the question arises as to whether our present
theorem can be recast in a form that is so explicitly independent of the
null generic condition. Preliminary investigations show that it can, but at
the cost of introducing new assumptions concerning the global structure
which serve to obscure somewhat the role of cosmic censorship. Whether a
more suitably direct theorem along these lines can be proved remains an
open issue.


\bigskip
\cent{\bf Acknowledgements}
\smallskip
\noindent
We would like to thank the Banach Center, Polish Academy of Sciences, and the
organizers of its Minisemester on Mathematical Aspects in Theories of
Gravitation for hospitality during the initiation of this work.


\bigskip
\cent{\bf References}
\smallskip
\item{1.}{J.L.~Friedman, K.~Schleich, and D.W.~Witt, {\it Topological
censorship}, Phys.~Rev. Lett.~71 (1993), 1486--1489.}
\item{2.}{G.~Galloway, {\it On the topology of the domain of outer
communication}, Class. Quantum Gravit.~12 (1995), L99--L101.}
\item{3.}{G.~Galloway, {\it A `finite infinity' version of topological
censorship}, Class. Quantum Gravit.~13 (1996), 1471--1478.}
\item{4.}{G.A.~Burnett, {\it Counterexample to the passive topological
censorship of $K(\pi,1)$ prime factors}, preprint 1995, available from
gr-qc/9504012.}
\item{5.}{P.T.~Chru\'sciel and R.M.~Wald, {\it On the topology of
stationary black holes}, Class. Quantum Gravit.~11 (1994), L147--L152.}
\item{6.}{T.~Jacobson and S.~Venkataramani, {\it Topology of event horizons
and topological censorship}, Class.~Quantum Gravit.~12 (1995), 1055--1061.}
\item{7.}{D.~Gannon, {\it Singularities in nonsimply connected
space-times}, J.~Math.~Phys.~16 (1975), 2364--2367.}
\item{8.}{R.~Penrose, {\it Singularities of spacetimes}, in: Theoretical
Principles in Astrophysics and Relativity, N.R. Lebovitz, W.H. Reid and P.O.
Vandervoort, eds., University of Chicago Press, Chicago,
1978, 217--243.}
\item{9.}{F.J.~Tipler,
{\it Energy conditions and spacetime singularities}, Phys.~Rev.~D17 (1978),
2521--2528; P.E. Ehrlich and S.-B. Kim, {\it From the Ricatti inequality to
the Raychaudhuri
equation}, in: Differential Geometry and Mathematical Physics, J.K. Beem
and K.L. Duggal, eds.,
Contemporary Math. Series, Amer. Math. Soc. 170 (1994), 65--78.}
\item{10.}{A.~Borde, {\it Geodesic focusing, energy conditions and
singularities}, Class. Quantum Gravit. 4 (1987), 343--356.}
\item{11.}{F.~Schein P.C.~and Aichelburg, {\it Traversable Wormholes in
Geometries of Charged Shells}, preprint 1996, available from
gr-qc/9606069.}
\item{12.}{S.W.~Hawking and G.F.R.~Ellis, {\it The large
scale structure of space-time}, Cambridge University Press, Cambridge,
1973, p.~265.}

%

\par\vfill\eject\bye